\def\deg{\ensuremath{^\circ}}
\def\sun{\hbox{$\odot$}}
\def\ga{\mathrel{\mathchoice {\vcenter{\offinterlineskip\halign{\hfil
$\displaystyle##$\hfil\cr>\cr\sim\cr}}}
{\vcenter{\offinterlineskip\halign{\hfil$\textstyle##$\hfil\cr
>\cr\sim\cr}}}
{\vcenter{\offinterlineskip\halign{\hfil$\scriptstyle##$\hfil\cr
>\cr\sim\cr}}}
{\vcenter{\offinterlineskip\halign{\hfil$\scriptscriptstyle##$\hfil\cr
>\cr\sim\cr}}}}}
\newcommand{\pcmq}{\mbox{cm$^{-2}$}}
\newcommand{\psec}{\mbox{s$^{-1}$}}

\newcommand{\funit}{\mbox{ph~\pcmq~\psec}}

\newcommand{\eunit}{\mbox{erg~\pcmq~\psec}}

\newcommand{\ra}{\mbox{$\alpha_{\rm J2000}$}}
\newcommand{\dec}{\mbox{$\delta_{\rm J2000}$}}
\newcommand{\hi}{\mbox{H\,{\scriptsize I}}}
\newcommand{\hii}{\mbox{H\,{\scriptsize II}}}

\newcommand{\hmol}{\mbox{H$_2$}}

\newcommand{\Msol}{\mbox{$M_{\sun}$}}

\newcommand{\ecut}{\mbox{$E_{\rm c}$}}
\newcommand{\psra}{\mbox{PSR~J0540$-$6919}}
\newcommand{\psrb}{\mbox{PSR~J0537$-$6910}}

\documentclass[twocolumn,twoside,slac_two]{revtex4}
\usepackage{graphicx}
\usepackage{fancyhdr}
\begin{document}

\title{Observations of the Large Magellanic Cloud with Fermi}

%

\author{J. Kn\"odlseder}
\affiliation{Centre d'\'Etude Spatiale des Rayonnements, CNRS/UPS, BP 44346, F-31028 Toulouse Cedex 4, France}
\author{P. Jean}
\affiliation{Centre d'\'Etude Spatiale des Rayonnements, CNRS/UPS, BP 44346, F-31028 Toulouse Cedex 4, France}
\author{on behalf of the Fermi/LAT Collaboration}

\begin{abstract}
We report on observations of the Large Magellanic Cloud with the Fermi 
Gamma-Ray Space Telescope. 
The LMC is clearly detected with the Large Area Telescope (LAT) and for 
the first time the emission is spatially well resolved in gamma-rays. 
Our observations reveal the massive star forming region 30 Doradus as 
a bright source of gamma-ray emission in the LMC. 
The observations furthermore show that the gamma-ray emission correlates 
little with the gas density of the LMC. 
Implications of this finding will be discussed.
\end{abstract}

\maketitle

\thispagestyle{fancy}


\section{Introduction}

Since the early days of high-energy gamma-ray astronomy it has been clear that the
gamma-ray flux received at Earth is dominated by emission from the Galactic disk
\cite{clark68}.
This emission is believed to arise from cosmic-ray interactions with the interstellar medium, 
which at gamma-ray energies $\ga100$~MeV are dominated by the decay of $\pi^0$ 
produced in collisions between cosmic-ray nuclei and the interstellar medium \cite{pollack63}.
Further contributions are from cosmic-ray electrons undergoing inverse Compton scattering
off interstellar soft photons and Bremsstrahlung losses within the interstellar medium.
Gamma-ray observations thus have the potential to map cosmic-ray acceleration sites in our 
Galaxy which may ultimately help to identify the sources of cosmic-ray acceleration.

Nearby galaxies have the advantage of being viewed from outside and so line of sight confusion, 
which complicates studies of emission from the Galactic disk, is diminished.
This advantage is however somewhat offset by the limitations by the angular resolution 
and sensitivity of the instrument.
The Large Magellanic Cloud (LMC) is thus an excellent target for studying the link between 
cosmic-ray acceleration and gamma-ray emission since 
the galaxy is nearby ($D\approx50$~kpc) \cite{matsunaga09,pietrzynski09},
has a large angular extent of $\sim8\deg$,
and is seen at a small inclination angle of $i\approx20\deg-35\deg$
\cite{kim98,vandermarel06}
that avoids source confusion.
In addition, the LMC is relatively active, housing many supernova remnants, bubbles
and superbubbles, and massive star forming regions that are all potential sites
of cosmic-ray acceleration \cite{cesarsky83,biermann04,binns07}.

The EGRET telescope onboard the Compton Gamma-Ray Observatory 
({\em CGRO}, 1991--2000) was the first to detect the LMC \cite{sreekumar92}.
Due to EGRET's limited angular resolution and limited sensitivity, details of the
spatial structure of the gamma-ray emission could not be resolved, yet the 
observations showed some evidence for the spatial distribution being consistent
with the morphology of radio emission.
The Large Area Telescope (LAT) aboard {\em Fermi} is providing now for the first time 
the capabilities to go well beyond the study of the integrated gamma-ray flux from the 
LMC \cite{digel00,weidenspointner07}.
In this contribution we present our first in-depth analysis of the LMC galaxy based on 
11 months of continuous sky survey observations performed with {\em Fermi}/LAT.
We put a particular emphasis on the determination of the spatial distribution of the
gamma-ray emission, that, as we will show, reveals the distribution of cosmic rays
in the galaxy.
A more detailed discussion of the observations and their implications
is given in \cite{abdo10}.

\section{Observations}
\label{sec:observation}

\subsection{Data preparation}

\begin{figure*}[t]
\centering
\includegraphics[width=135mm]{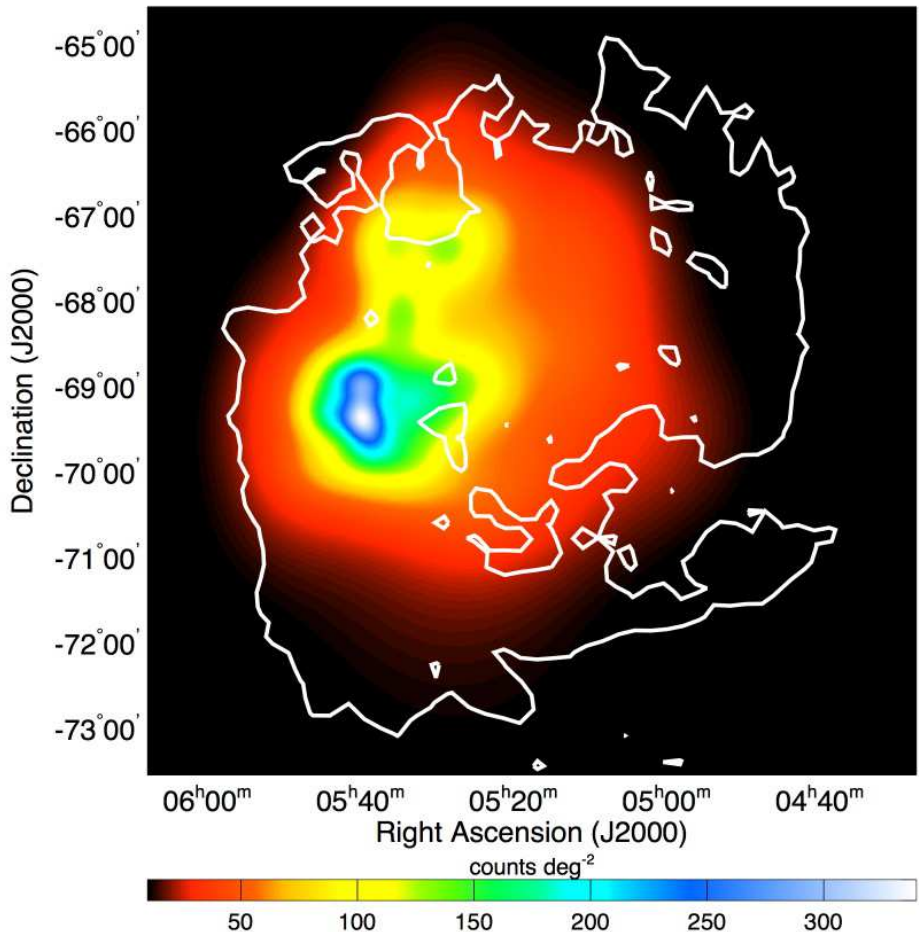}
\caption{Background subtracted counts map of the LMC region. 
The counts map has been adaptively smoothed \cite{ebeling06} with a signal-to-noise 
ratio of 5 in order to reveal significant structures at all possible scales
while supressing the noise that arises from photon counting statistics.
The white line shows
the N(\hi)~$=10^{21}$ H~cm$^{-2}$ iso column density contour to indicate the
extent of the gaseous disk.} \label{fig:map}
\end{figure*}

The characteristics and performance of the LAT aboard {\em Fermi}
are described in detail by \cite{atwood09}.
The data used in this work amount to 274.3 days of continuous sky survey observations
over the period
August 8th 2008 -- July 9th 2009
during which a total exposure of 
$\sim2.5 \times 10^{10}$~cm$^2$~s (at 1 GeV)
is obtained for the LMC.
Events satisfying the standard low-background event
selection (``Diffuse'' events) \cite{atwood09} and coming from zenith angles
$<105\deg$ (to greatly reduce the contribution by Earth albedo gamma rays)
are used.
To further reduce the effect of Earth albedo backgrounds, the time intervals
when the Earth was appreciably within the field of view (specifically, when
the center of the field of view was more than $47\deg$ from the zenith)
are excluded from this analysis.
Furthermore,  time intervals when the spacecraft was within the South Atlantic Anomaly 
are also excluded.
We further restrict the analysis to photon energies above 200~MeV; below this energy
the effective area in the ``Diffuse class" is relatively small and strongly dependent on 
energy.
All analysis is performed using the LAT Science Tools package, which is available 
from the Fermi Science Support Center, using P6\_V3 post-launch instrument response 
functions (IRFs).
These take into account pile-up and accidental coincidence effects in the
detector subsystems that are not considered in the definition of the pre-launch
IRFs.

At the Galactic latitude of the LMC ($b \approx -33\deg$), the gamma-ray background is
a combination of extragalactic and Galactic diffuse emissions,
some residual instrumental background and a number of point sources that primarily
are associated to blazars.
We model the background using components for the diffuse Galactic and 
the extragalactic and residual instrumental backgrounds and 
6 point sources that all are associated with known blazars.
The Galactic component is based on the LAT standard diffuse background model
{\tt gll\_iem\_v02}
for which we keep the overall normalization as a free parameter.
The extragalactic and residual instrumental backgrounds are combined into a single
component which has been taken as being isotropic.
The spectrum of this component is determined by fitting an isotropic component
together with a model of the Galactic diffuse emission and point sources to the data.
Also here we leave the overall normalization of the component as a free parameter.
The 6 background blazars are modelled as point sources with power-law spectral shapes.
The flux and spectral power-law index of each source are left as free parameters of our 
background model and their values were determined from likelihood analysis.

\subsection{Spatial distribution}
\label{sec:spatial}

\subsubsection{Counts map}

\begin{figure}[t]
\centering
\includegraphics[width=90mm]{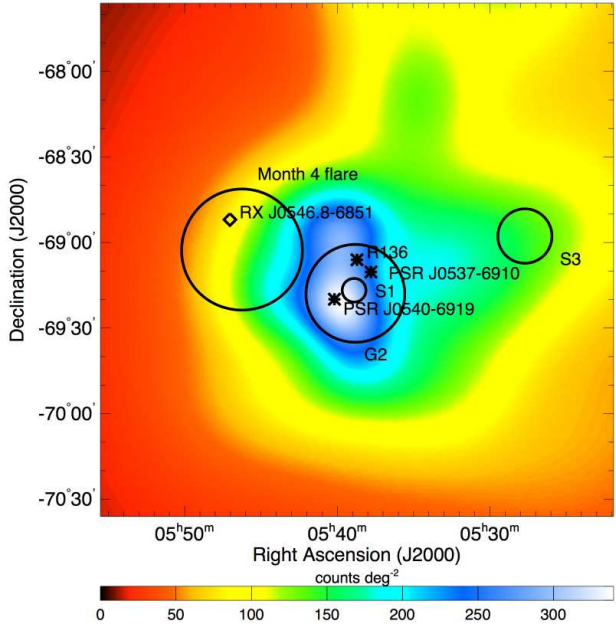}
\caption{Zoom into a $3\deg \times 3\deg$ large region of the background subtracted counts map 
around the central star cluster R~136 in 30~Dor.
Stars show the locations of R~136 and of the pulsars \psra\ and \psrb.
The circles show the 95\% containment radius of sources S1 and S3 of the
point source model, and of source G2 of the 2D Gaussian shaped model
(see text).
During month 4 of our dataset, a flaring point source occured near 30~Dor,
and we indicate the 95\% containment radius of this source; the diamond shows 
the location of the possible counterpart RX~J0546.8$-$6851 of the flaring source.} 
\label{fig:30Dor}
\end{figure}

To investigate the spatial distribution of gamma-ray emission toward the LMC we show
in Fig.~\ref{fig:map} an adaptively smoothed background subtracted counts map of the
LAT data.
The map clearly shows extended emission that is spatially confined to within the LMC
boundaries which we trace by the iso column density contour 
$N_{\rm H} = 10^{21}$ H~cm$^{-2}$
of neutral hydrogen in the LMC \cite{kim05}.
The total number of excess 200 MeV -- 20 GeV photons above the background in the LMC 
area amounts to $\sim1550$ counts whereas the background in the same area amounts to 
$\sim2440$ counts.
With these statistics, the extended gamma-ray emission from the LMC can be
resolved into several components.
The brightest emission feature is located near
$(\ra, \dec) \approx (05^{\rm h}40^{\rm m}, -69\deg15')$,
which is close to the massive star-forming region 30~Doradus (30~Dor)
that houses the two Crab-like pulsars 
\psra\ and \psrb\ \cite{seward84,marschall98}.
Excess gamma-ray emission is also seen toward the north and the west of 30~Dor.
These bright regions are embedded into a more extended and diffuse glow that
covers an area of approximately $5\deg \times 5\deg$.
Figure~\ref{fig:30Dor} present a zoom of Fig.~\ref{fig:map} of the 30~Dor region over
which we overlay potential sources of gamma-ray emission.

\subsubsection{Model fitting}
\label{sec:models}

As next step, we assess the spatial distribution of the LMC emission using 
(1) simple parametrized geometrical models of the gamma-ray intensity distribution, and
(2) spatial templates that trace the interstellar matter distribution in the LMC.
We assume power-law spectral distributions for all models and keep the 
total flux and power law index as free parameters.
We adjust the spatial and spectral parameters of the models using a binned maximum 
likelihood analysis with spatial bins of $0.1\deg \times 0.1\deg$ and 60 logarithmically 
spaced energy bins covering the energy range 200 MeV -- 20 GeV.
We quantify the goodness-of-fit using the so-called {\em Test Statistic} (TS) which is defined as 
twice the difference between 
the log-likelihood $\mathcal{L}_1$ that is obtained by fitting 
the model on top of the background model to the data, and
the log-likelihood $\mathcal{L}_0$ that is obtained by fitting the background model only,
i.e. ${\rm TS} = 2(\mathcal{L}_1 - \mathcal{L}_0)$.

\begin{figure*}[t]
\centering
\includegraphics[width=56mm]{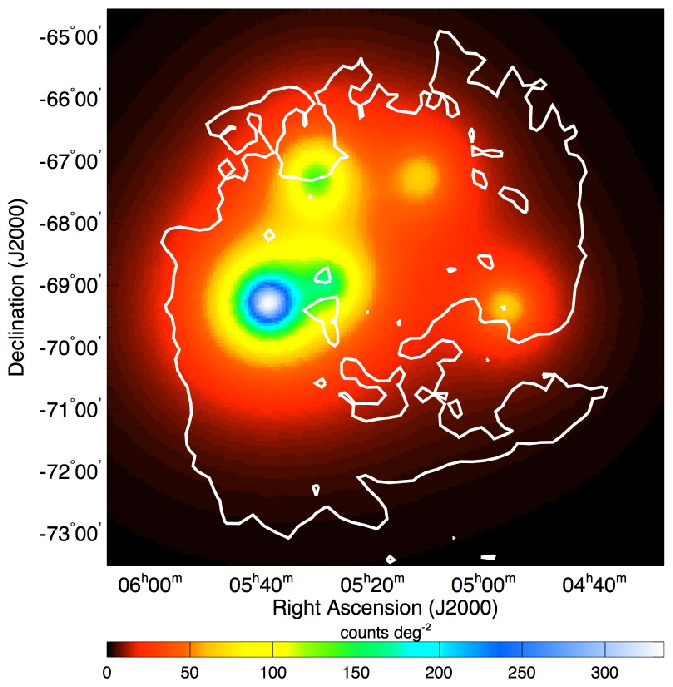}
\includegraphics[width=56mm]{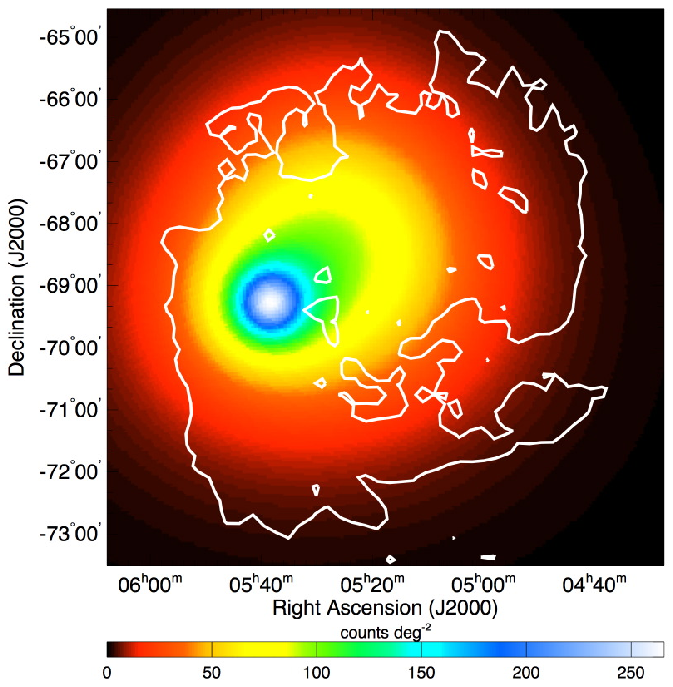}
\caption{Best fitting point source model (left panel) and 2D Gaussian shape 
model (right panel) that has been convolved with the LAT point spread function.
The point source model consists of 5 point sources (S1-S5) while the 2D Gaussian
shaped model consists of a broad component covering a large fraction of the
LMC disk (G1) and a narrow component near 30~Dor (G2).} 
\label{fig:geo}
\end{figure*}

First, we examine if the gamma-ray emission from the LMC can be explained with a
combination of individual point sources.
For this purpose we add successive point sources to our model and optimize their
locations, fluxes and spectral indices by maximizing the likelihood of the model.
We stop this procedure once the TS improvement after adding a further point source
drops below 25.
This happens after we added 5 point source to our model, resulting in TS~$=1089.3$.
The left panel of Fig.~\ref{fig:geo} shows the corresponding model counts map.

Second, instead of using point sources we repeat the procedure with 2D Gaussian shaped
intensity profiles to build a geometrical model that is more appropriate for extended and 
diffuse emission structures.
We again stop the successive addition of 2D Gaussian shaped sources once the TS
improvement after adding a further source drops below 25.
This occurs after two 2D Gaussian shaped sources have been added to the model,
resulting in TS~$=1122.6$.
We show the corresponding model counts map in the right panel of Fig.~\ref{fig:geo}.
Obviously, the 2D Gaussian model provides a larger TS than the point source model,
suggesting that it better fits the data despite the smaller number of
free parameter (10 for the 2D Gaussian model compared to 20 for the point source model).
It is thus more likely that the LMC emission is indeed diffuse in nature, or alternatively,
composed of a large number of unresolved and faint sources that can not be detected
individually by {\em Fermi}/LAT.

\begin{figure*}[t]
\centering
\includegraphics[width=56mm]{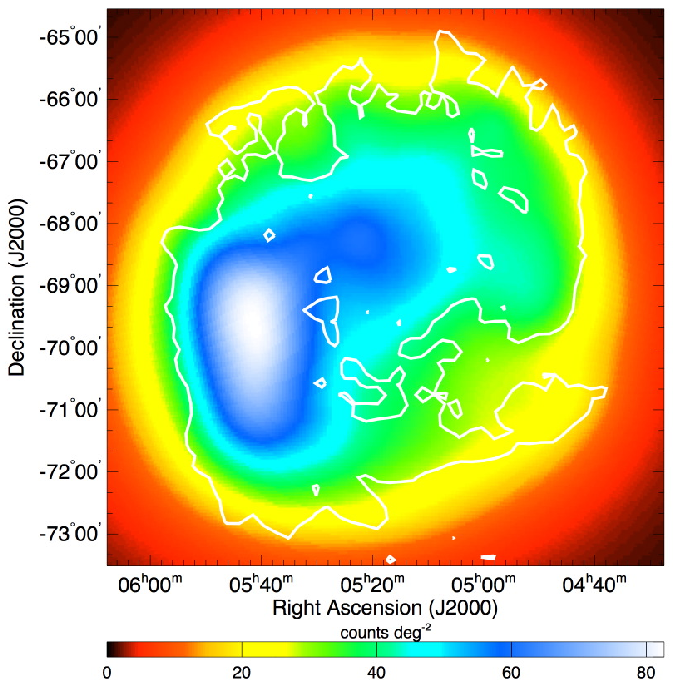}
\includegraphics[width=56mm]{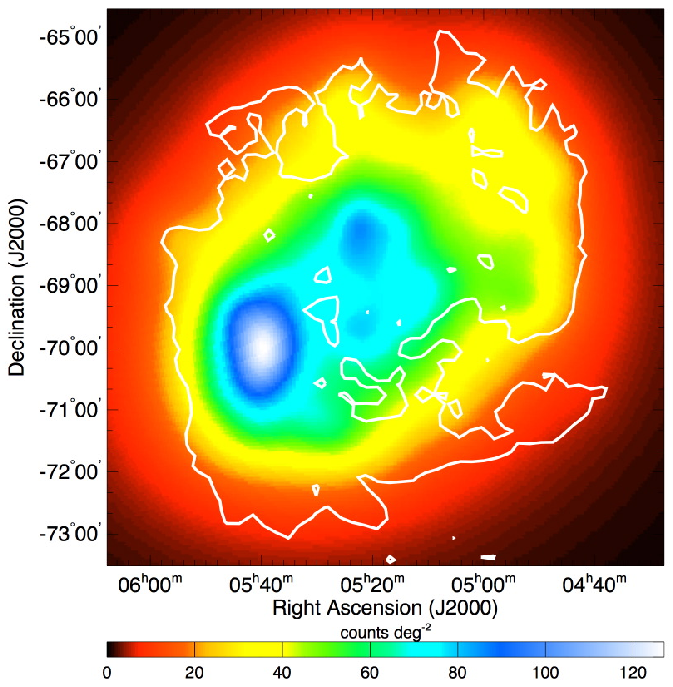}
\includegraphics[width=56mm]{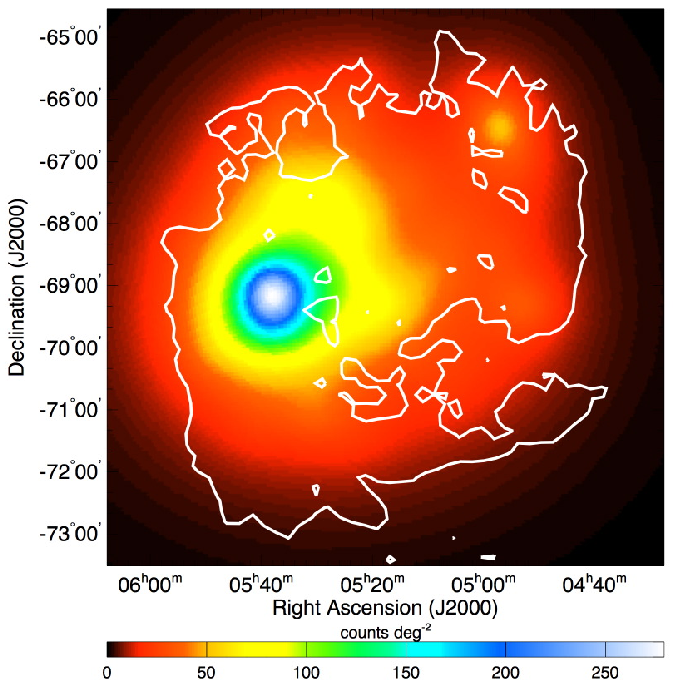}
\caption{Gas tracer maps convolved with the LAT point spread function and scaled using
maximum likelihood model fitting to the data. Panels show
\hi\ (left panel),
\hmol\ (mid panel), and
\hii\ (right panel).} \label{fig:tracer}
\end{figure*}

Third, we fit the LAT data to gas maps of 
neutral atomic hydrogen (\hi) \cite{kim05},
molecular hydrogen (\hmol) \cite{fukui08}, and
ionized hydrogen (\hii) \cite{finkbeiner03}.
Fitting the \hi\ and \hmol\ maps result in TS values of $771.8$ and $824.3$, respectively,
that are considerably worse than those obtained for the geometrical models.
Apparently, the \hi\ and \hmol\ maps provide rather poor fits to the data, indicating 
that the distribution of gamma rays does not follow the distribution of neutral hydrogen 
in the LMC.
This can already be seen from the left and mid panels of Fig.~\ref{fig:tracer} which show
the gas maps after convolution with the LAT point spread function.
While the gamma-ray emissivity is highest in 30~Dor and the northern part of the galaxy, 
the gas maps show a bright ridge that runs over $\sim3\deg$ along 
$\ra\sim05^{\rm h}40^{\rm m}$ which coincides with the most prominent region of
$^{12}$CO emission tracing giant molecular clouds in the LMC \cite{fukui99}.
Roughly $20\%$ of the total gas mass in the LMC is confined into this ridge \cite{luks92}, 
yet comparison with Fig.~\ref{fig:map} shows that most of the ridge is not luminuous in 
high-energy gamma rays.

The \hii\ map, on the other hand, gives TS~$=1110.1$, which is very close to the TS values
of the geometrical models.
The \hii\ map thus provides the best fit among all of the gas maps to the LAT data, and
the right panel of Fig.~\ref{fig:tracer} indeed shows that the model is very similar to the
2D Gaussian shaped model and also follows closely the observed distribution of gamma
rays (Fig.~\ref{fig:map}).

\subsection{Emissivity spectrum}

To determine the spectrum of the gamma-ray emission 
from the LMC independently from any assumption on the spectral shape, we fit our data in  
6 logarithmically spaced energy bins covering the energy range 200 MeV - 20 GeV.
We obtain the total spectrum of the LMC by fitting the \hii\ template to the data.
We also obtain separate spectra for the LMC disk (G1) and for 30~Dor (G2) by fitting 
the 2D Gaussian shaped model to the data.

To determine the integrated gamma-ray flux we fit exponentially cut off power 
law spectral models of the form
$N(E) = k\,(E/E_0)^{-\Gamma} \exp(-E/\ecut)$
to the data.
We make these fits by means of a binned maximum likelihood analysis over the energy range 
200 MeV - 20 GeV.
This results in an extrapolated $>100$~MeV photons flux of
$(2.6 \pm 0.2) \times 10^{-7}$ \funit\
for the \hii\ template which corresponds to an energy flux of
$(1.6 \pm 0.1) \times 10^{-10}$ \eunit\
(systematic uncertainties in these estimates amount to less than 16\%).

\begin{figure}[t]
\centering
\includegraphics[width=82mm]{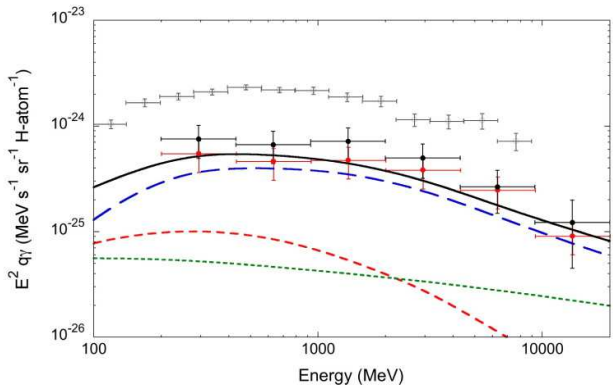}
\caption{Differential average gamma-ray emissivity for the total LMC emission
(black dots) and for the LMC disk only (red dots) 
compared to that of the Galactic local interstellar medium determined by 
\cite{abdo09} (grey data points).
The solid line shows the predicted gamma-ray emissivity computed in the framework
of a one-zone model for the LMC disk \cite{abdo10}.
The other lines show the contributions of 
$\pi^0$-decay (long dashed),
Bremsstrahlung (short dashed), and
inverse Compton emission (dotted).} \label{fig:spectrum}
\end{figure}

Dividing our spectra by the spatially integrated hydrogen column density
$\int N_{\rm H}~{\rm d}\Omega = (3.6 \pm 1.2) \times 10^{19}$~H-atom~cm$^{-2}$~sr
of the LMC provides us with the differential gamma-ray emissivity per hydrogen
atom.
We compare the resulting emissivity for the total LMC emission and the LMC disk
only in Fig.~\ref{fig:spectrum}
to the one obtained for the Galactic local interstellar medium by 
\cite{abdo09}.
Apparently, the local Galactic emissivity is between $\sim2$ to $\sim4$ times larger than
the average emissivity of the LMC.

We compare the differential gamma-ray emissivities to a one-zone model of 
cosmic-ray interactions with the interstellar medium that takes into account $\pi^0$ 
decay following proton-proton interactions, Bremsstrahlung from
cosmic-ray electrons and inverse Compton scattering of cosmic-ray electrons on LMC
optical and infrared photons and cosmic microwave background photons
\cite{abdo10}.
Fitting this spectral model to our data using a binned maximum likelihood analysis gives an 
average cosmic-ray enhancement factor of 
$r_c = 0.31 \pm 0.01$ for the entire LMC, and of
$r_c = 0.21 \pm 0.01$ for the LMC disk only.
Systematic errors due to uncertainties in the effective area of the instrument amount to
$\pm0.02$.
An additional systematic error of $-23\%$ to $+42\%$ comes from the uncertainty in the 
total gas mass of the LMC, which largely dominates the statistical and systematic
measurements errors.

\section{Cosmic-ray density distribution}

\begin{figure*}[t]
\centering
\includegraphics[width=135mm]{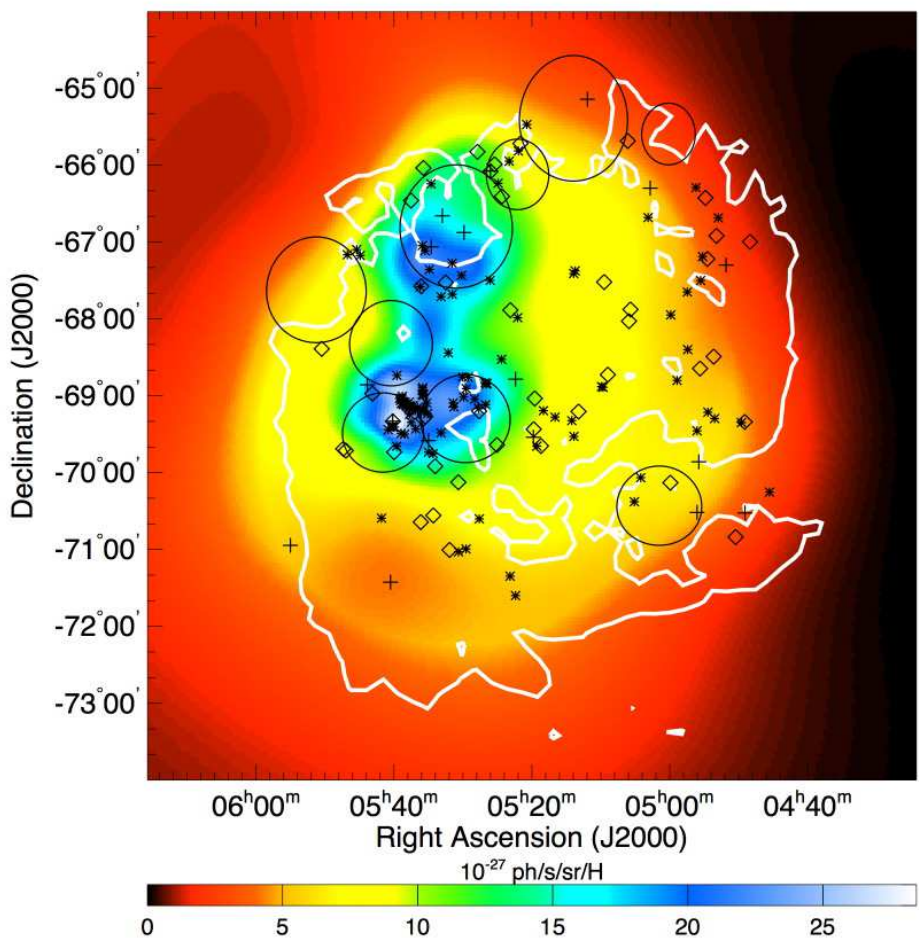}
\caption{Integrated $>100$~MeV emissivity maps of the LMC in units of
$10^{-27}$~ph~s$^{-1}$~sr$^{-1}$~H-atom$^{-1}$.
An adaptive smoothing with a signal-to-noise ratio of 5 has been applied to reduce statistical
fluctuations.
The white line shows
the N(\hi)~$=10^{21}$ H~cm$^{-2}$ iso column density contour to indicate the
extent of the gaseous disk.
Symbols indicate the locations of 
pulsars (pluses) from ATNF catalogue version 1.36 \cite{manchester05},
supernova remnants (diamonds) from Rosa Williams web page 
http://www.astro.illinois.edu/projects/atlas/index.html,
Wolf-Rayet stars (stars) from the fourth catalogue of \cite{breysacher99},
and supergiant shells (circles) from \cite{staveleysmith03}.} \label{fig:emissivitymap}
\end{figure*}

To reveal the sites of cosmic-ray acceleration in the LMC we map the cosmic-ray density 
variations  in the galaxy by computing the gamma-ray emissivity as function 
of position.
We do this by dividing our background subtracted counts map by the 
$N({\rm H})$ map after convolution of the latter with the LAT instrumental response function.
We normalise $N({\rm H})$ to a total LMC hydrogen mass of 
$7.2 \times 10^8$ \Msol\
that takes into account the possible presence of dark gas that is not seen in radio surveys
of \hi\ \cite{bernard08}.
We adaptively smooth \cite{ebeling06} the counts maps and used the resulting smoothing
kernel distribution to smooth also the convolved $N({\rm H})$ map before the division to 
reveal significant structures at all possible scales, while supressing the noise that arises 
from the limited photon counting statistics.
The resulting emissivity maps are shown in Fig.~\ref{fig:emissivitymap}.
We superimpose on the images the interstellar gas distribution, as traced by $N({\rm H})$, 
convolved with the LAT instrumental response function, and also show the locations of 
potential particle acceleration sites, such as pulsars, supernova remnants, Wolf-Rayet stars 
and supergiant shells.

Figure~\ref{fig:emissivitymap} reveals that the cosmic-ray density varies considerably
over the disk of the LMC.
The gamma-ray emissivity is highest in 30~Dor and the northern part of the galaxy, while 
the southern part and in particular the dense ridge of gas south of 30~Dor seems basically
devoid of cosmic rays.
These large variations confirm our earlier findings that the gamma-ray emission correlates 
little with the gas density in the LMC.
Figure~\ref{fig:emissivitymap} suggests further that the cosmic-ray density correlates
with massive star forming tracers, and in particular Wolf-Rayet stars and supergiant 
shells.
This finding is corroborated by the good fit of the \hii\ gas map, which is probably the most
direct tracer of massive star forming regions within a galaxy.

Thus, the gamma-ray emissivity maps of the LMC support the idea that cosmic rays
are accelerated in massive star forming regions as a result of the large amounts of kinetic
energy that are input by the stellar winds and supernova explosions of massive stars into
the interstellar medium.
Our data reveal a relatively tight confinement of the gamma-ray emission
to star forming regions, which suggests a relatively short diffusion length for GeV
protons.

\section{Conclusions}
\label{sec:conclusion}

Observations of the LMC by {\em Fermi}/LAT have for the first time provided a detailed
map of high-energy gamma-ray emission from that galaxy.
Our analysis reveals the massive star forming region 30~Doradus as bright source of
gamma-ray emission in the LMC in addition to fainter emission regions found in the northern
part of the galaxy.
The gamma-ray emission from the LMC shows very little correlation
with gas density.
A much better correlation is seen between gamma-ray emission and massive star
forming regions, as traced by the ionizing gas, Wolf-Rayet stars and supergiant shells,
and we take this as evidence for cosmic-ray acceleration in these regions.
This correlation supports the idea that cosmic rays are accelerated in massive star forming 
regions as a result of the large amounts of kinetic energy that are input by the stellar winds 
and supernova explosions of massive stars into the interstellar medium.

Continuing observations of the LMC with {\em Fermi}/LAT in the upcoming
years will provide the photon statistics to learn more about the origin of the gamma-ray
emission from that galaxy.
Better statistics will help in identifying more individual emission components and may help
to separate true point sources from the more diffuse emission that we expect from
cosmic-ray interactions.

\bigskip 
\begin{acknowledgments}
The \textit{Fermi} LAT Collaboration acknowledges generous ongoing support
from a number of agencies and institutes that have supported both the
development and the operation of the LAT as well as scientific data analysis.
These include the National Aeronautics and Space Administration and the 
Department of Energy in the United States, the Commissariat \`a l'Energie Atomique
and the Centre National de la Recherche Scientifique / Institut National de Physique
Nucl\'eaire et de Physique des Particules in France, the Agenzia Spaziale Italiana
and the Istituto Nazionale di Fisica Nucleare in Italy, the Ministry of Education,
Culture, Sports, Science and Technology (MEXT), High Energy Accelerator Research
Organization (KEK) and Japan Aerospace Exploration Agency (JAXA) in Japan, and
the K.~A.~Wallenberg Foundation, the Swedish Research Council and the
Swedish National Space Board in Sweden.

Additional support for science analysis during the operations phase is gratefully
acknowledged from the Istituto Nazionale di Astrofisica in Italy and the and the Centre National d'\'Etudes Spatiales in France.
\end{acknowledgments}

\bigskip 

\end{document}